\documentclass[epj]{svjour}
\usepackage{amsmath}
\usepackage[T1]{fontenc} 
\usepackage[mathlines,displaymath]{lineno}
\usepackage{graphics}
\newcommand{\pt}{\mathrm{p_T}}
\begin{document}
\title{ABCNet: An attention-based method for particle tagging}
\author{V. Mikuni\inst{1} \and F. Canelli\inst{1}
}                     
\offprints{}          
\institute{University of Zurich Winterthurerstrasse 190. CH-8057 Zurich.}
%
%
\abstract{
In high energy physics, graph-based implementations have the advantage of treating the input data sets in a similar way as they are collected by collider experiments. To expand on this concept, we propose a graph neural network enhanced by attention mechanisms called ABCNet. To exemplify the advantages and flexibility of treating collider data as a point cloud, two physically motivated problems are investigated: quark-gluon discrimination and pileup reduction. The former is an event-by-event classification while the latter requires each reconstructed particle to receive a classification score. For both tasks ABCNet shows an improved performance compared to other algorithms available.
\PACS{
      {PACS-key}{discribing text of that key}   \and
      {PACS-key}{discribing text of that key}
     } 
} 
\maketitle
\section{Introduction}
\label{sec:intro}
One of the main goals in modern machine learning is to be able to extract the maximum amount of information available from a data set. Successful implementations take advantage of the data structure for model building. In high energy physics (HEP), particle collisions in experiments are reconstructed by combining the energy deposits left by particles after crossing different parts of a detector. The information provided by sub-detectors can be further combined to give a full description of each particle produced. At the Large Hadron Collider (LHC) \cite{Evans:2008zzb}, jets are ubiquitous objects produced in proton-proton collisions. Jets are the byproducts of the hadronisation of quarks and gluons, resulting in an often collimated spray of particles. After each collision, $\mathcal{O}(1000)$ or more particles can be produced, making the task of identifying the original hard scattering objects challenging.
The luminosity increase at the LHC will also increase the amount of multiple interactions per bunch crossing (pileup). For instance, event collisions recorded thus far by the ATLAS \cite{Aad:2008zzm} and CMS \cite{Chatrchyan:2008aa} detectors at LHC measured an average of about 30 extraneous interactions. With the future upgrade, up to 200 pileup events per bunch crossing are expected, requiring new methods for particle identification and pileup suppression. In this paper a new method for event classification in HEP is introduced. The attention-based cloud net (ABCNet) takes into account the data structure recorded by particle collision experiments, treating each interaction as an unordered set of points that defines a point cloud. This description is advantageous since the byproducts of each particle collision are treated in a similar fashion as they are collected by particle detectors. To enhance the extraction of local information, an attention mechanism is used, following closely the implementation developed in \cite{2019arXiv190508705C}.
Attention mechanisms have proved to boost performance for different applications in machine learning by giving local and global context to the learning procedure. 
To show the performance and flexibility of the model, two critical problems are investigated: quark-gluon discrimination and pileup mitigation. 
  
\section{Related works}
\label{sec:related}
The main novelties introduced by ABCNet are the treatment of particle collision data as a set of permutation invariant objects, enhanced by attention mechanisms to filter out the particles that are not relevant for the tasks we want to accomplish.
The usage of graph-based machine learning implementations is still a new concept in particle physics. Nevertheless, new implementations have already been proposed with promising results. ParticleNet \cite{2019arXiv190208570Q} uses a similar approach, using point clouds for jet identification. The main difference between ABCNet and ParticleNet is that ABCNet takes advantage of attention mechanisms to enhance the local feature extraction, allowing for a more compact and efficient architecture. A theory-inspired approach was also developed in the framework of Deep Sets \cite{NIPS2017_6931} using an infrared and collinear safe basis, developed in the context of Energy Flow Networks \cite{Komiske_2019}. A message-passing approach for jet-tagging discussed in \cite{bib:message_passing}. Interaction networks were also studied in the context of high-mass particle decays with JEDI-net \cite{Moreno:2019bmu}. Other graph-based implementations have also been presented in the context of signal and background classification \cite{Abdughani:2018wrw,DBLP:journals/corr/abs-1809-06166}, particle track reconstruction \cite{osti_1484458}, and particle reconstruction  on irregular calorimeters. \cite{Qasim2019}. In the context of pileup rejection, the GGNN implementation \cite{ArjonaMartinez2019} shows promising results by combining graph nodes with GRU cells. 

\section{GAPLayer}
\label{sec:net}
ABCNet follows closely the implementation described for GAPNet \cite{2019arXiv190508705C},  with key differences to adapt the implementation to our problems of interest. For clarity, the description of the essential aspects of the implementation are described. The key aspect of GAPNet is the development of a graph attention pooling layer (GAPLayer) using the edge convolution operation proposed in \cite{DBLP:journals/corr/abs-1801-07829}, which defines a convolution-like operation on point clouds together with attention mechanisms to operate on graph-structured data described in \cite{velikovi2017graph}. The point cloud is first represented as a graph with vertices represented by the points themselves. The edges are constructed by connecting the points to their k-nearest neighbours, while the edge features, $y_{ij} = (x_i - x_{ij})$, are taken as the difference between features of each point $x_i$ and its k-neighbours $x_{ij}$. 
A GAPLayer is constructed by first encoding each point and edge to a higher-level feature space of dimension F using a single-layer neural network (NN), with learnable parameters $\theta$, in the following form:
\begin{align*}
    x_i'& = h(x_i,\theta_i, F)      \\
    y_{ij}'& = h(y_{ij},\theta_{ij}, F)      
\end{align*}
Where h() denotes the single-layer neural network operation. Self- and local-coefficients are created by passing the transformed points and edges to a single-layer NN with output dimension of size one. Finally, the attention coefficients $c_{ij}$  are created by combining the newly created coefficients in the following way:

\begin{linenomath}
\begin{equation}
    c_{ij} = \mathrm{LeakyRelu}(h(x'_i,\theta_i', 1) +  h(y_{ij}',\theta_{ij}', 1))
\end{equation}
\end{linenomath}

where the non-linear LeakyRelu operation is applied to the output of the sum. To align the attention coefficients between different points, a Softmax normalisation is applied to the coefficients $c_{ij}$. 
At this moment, each point is associated to \textit{k} attention coefficients. To compute a single attention feature for each point, a linear combination with a non-linear activation function is defined as
\begin{linenomath}
\begin{equation}
    \hat{x}_i = \mathrm{Relu}\left ( \sum_j c_{ij}y'_{ij} \right).
\end{equation}
\end{linenomath}
To enhance the stability of the determination of the coefficients $\hat{x}_i$, a multi-head mechanism can be used. A M-head process repeats the same procedure described above, determining  $\hat{x}_i$ M times, differing only on the random weight initialisation. The M results are combined by taking the maximum of the M different $\hat{x}_i$ .
The outputs of each GAPLayer consist of attention features ($\hat{x}_i$) and graph features ($y'_{ij}$). The graph features are further aggregated in the form:

\begin{linenomath}
\begin{equation*}
    y_{ij}^{max} = max(y_{ij}').
\end{equation*}
\end{linenomath}
Due to stackability properties, a GAPlayer output can be further used as an input to a subsequent GAPLayer or multilayer perceptron (MLP).

\section{Classification: Quark-gluon tagging}
Quark-gluon tagging refers to the task of identifying the origin of a jet as produced from the hadronisation of a gluon or a quark. The data set used for the studies are available from \cite{Komiske_2019}. It consists of stable particles clustered into jets, excluding neutrinos, using the anti-$k_{T}$ algorithm \cite{Cacciari:2008gp} with R=0.4. The quark-initiated sample (signal) is generated using a Z($\nu\nu$) + $(u,d,s)$ while the gluon-initiated data (background) are generated using Z($\nu\nu$) +$g$ processes. Both samples are generated using \textsc{Pythia8} \cite{Sjostrand:2014zea} without detector effects. Jets are required to have transverse momentum $\pt \in [500,550]$ GeV and rapidity $|y|<1.7$ for the reconstruction. For the training, testing and evaluation of the method, the  recommended splitting is used with 1.6M/200k/200k events respectively. 
For every reconstructed jet, up to 200 constituents are saved. Each constituent contains the four momentum and the expected particles type (electron, muon, photon, or charged/neutral hadrons). A typical jet has $\mathcal{O}(10)$ to $\mathcal{O}(100)$ particles. To simplify the implementation, ABCNet uses the first 100 constituents sorted by $p_{T}$ from highest to lowest. If the jet has less than 100 constituents, the event is padded with zeros, if there are more than 100 constituents, the event is truncated.

To enhance the non-local information extraction, global features can also be added to ABCNet. The approach is similar to the one described in \cite{Baldi:2016fzo}, where global information is used to parameterise the network, improving the generalisation and performance as a function of the global parameters.

The features used to describe each constituent are listed in Table~\ref{tab:qg_vars}.
\begin{table}[h]
    \centering
    \caption{Description of each feature used to define a point in the point cloud implementation for quark-gluon classification. The latter two features are the global information added to the network }
    \label{tab:qg_vars}
	\begin{tabular}{lc}
    \hline\noalign{\smallskip} 
             Variable & Description  \\
            \hline
            $\Delta\eta$       &  \small{Difference between the pseudo-rapidity of the constituent and the  jet}\\  
            $\Delta\phi$       &  \small{Difference between the azimuthal angle of the constituent and the  jet}\\  
            $\log\pt$       &  \small{logarithm of the constituent's $\pt$ }\\  
            $\log\mathrm{E}$       &  \small{logarithm of the constituent's E }\\  
            $\log\frac{\pt}{\pt(\mathrm{jet})}$       &  \small{logarithm of the ratio between the constituent's $\pt$ and the  jet $\pt$}\\  
            $\log\frac{\mathrm{E}}{\mathrm{E}(\mathrm{jet})}$       &  \small{logarithm of the ratio between the constituent's E and the jet E}\\  
            $\Delta\mathrm{R}$       &  \small{Distance in the $\eta-\phi$ space between the constituent and the  jet}\\  
            PID       &  \small{particle type identifier as described in \cite{Tanabashi:2018oca}. }\\  
            \hline
            m(jet) & \small{Jet mass} \\
            $\pt$(jet) & \small{Jet transverse momentum}  \\
    \noalign{\smallskip}\hline  
	\end{tabular}
\end{table}

\subsection{Network architecture}

\begin{figure*}
    \centering
    \resizebox{0.75\textwidth}{!}{
      \includegraphics{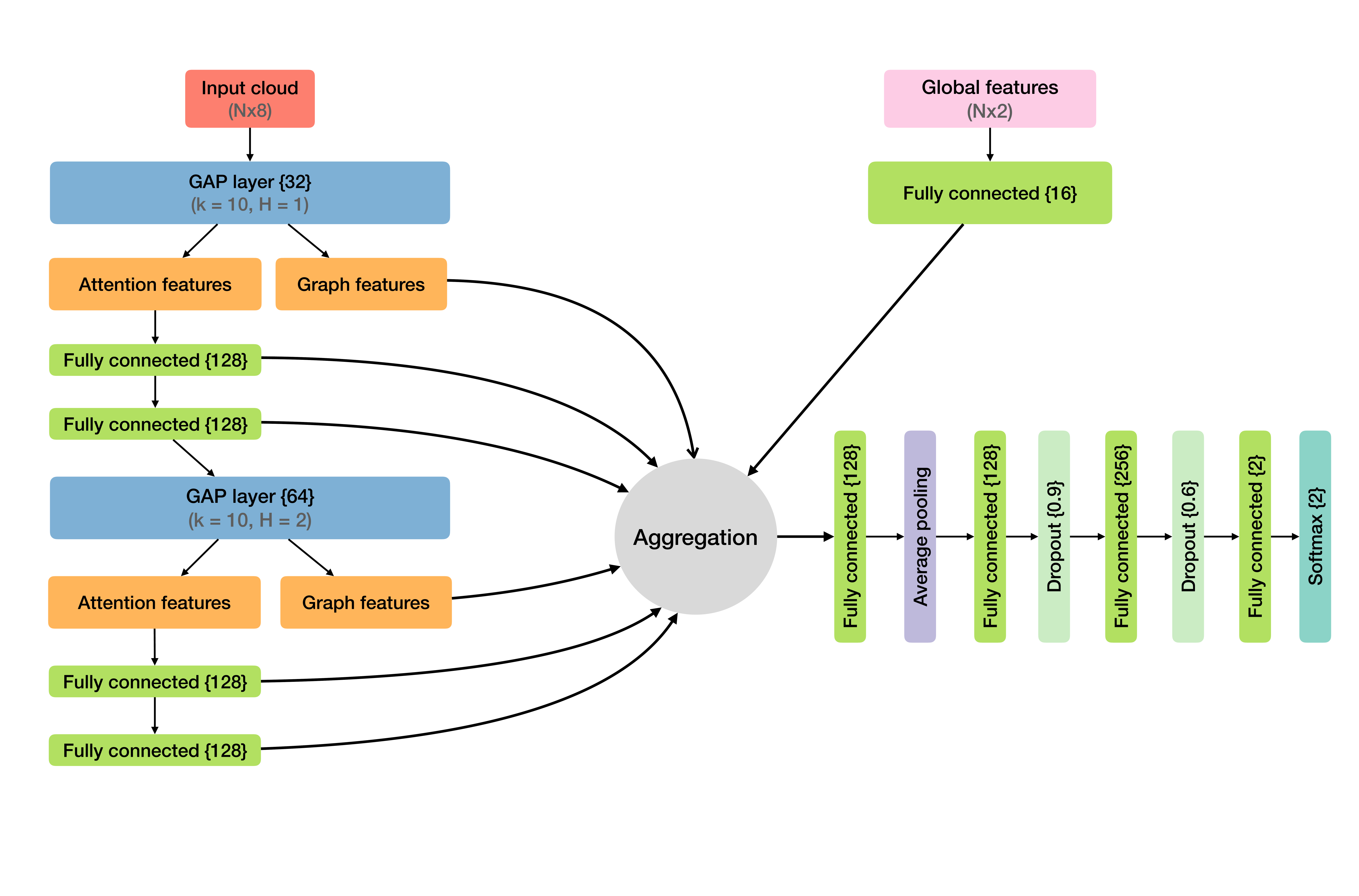}
    }   
    \caption{ABCNet architecture used for quark-gluon tagging. Fully connected layer and encoding node sizes are denoted inside ``\{\}''. For each GAPLayer, the number of k-nearest neighbours (k) and heads (H) are given.}
    \label{fig:ABC_qg}
\end{figure*}

The network layout used is shown in Fig.~\ref{fig:ABC_qg}. The first step is to calculate the distances between the constituents in the pseudorapidity-azimuth ($\eta-\phi$) space of the form $\Delta R = \sqrt{\Delta\eta^2 + \Delta\phi^2}$. From the distances, we create the first GAPLayer by associating each particle to its nearest 10 neighbours. While different choices for k were tested, the overall performance did not improve with the addition of more neighbours.  The encoding channel size of the GAPLayer F is selected to be 32 with a 1-head. The attention features created by the GAPLayer are then passed through two MLPs with node sizes (128,128). The distances used for the second GAPLayer are calculated using the full-feature space produced in the output of the last MLP, allowing the network to learn distances in the transformed feature space. To achieve a robust estimation, the encoding channel size is selected to be 64 with the number of heads determined to be two. The newly created attention features are passed through two MLPs of node sizes each of 128.
In parallel, ABCNet also takes additional global inputs in the form of the jet mass and transverse momenta. The global inputs are first transformed by means of a single-layer MLP with small node size of 16.
The two graph features and the output of each MLP are concatenated with the transformed global features and fed to a MLP of node size 128. An average pulling is applied and the result is further passed to 2 additional  MLPs of node sizes (128,256) interleaved by two dropout layers. A Softmax operation is applied to the output result.

\subsection{Results}
The performance of ABCNet is compared to the methods implemented in \cite{2019arXiv190208570Q} and \cite{Komiske_2019}, using the same data set. 

The figures of merit used for the comparison are:
\begin{itemize}
    \item Accuracy: Ratio between the number of correct predictions over the total number of test examples.
    \item AUC: Integral of the area under the receiver operating characteristic distribution.
    \item  1/$\epsilon_B$: One over the background efficiency for a fixed value of the signal efficiency (50\% or 30\%)
    \item Parameters: Number of trainable weights for the model.
\end{itemize}
The results of the comparisons are listed in Table~\ref{tab:qg_results}. Even though the accuracy obtained by ABCNet is numerically the same as the one reported by ParticleNet, ABCNet excels on the other figures of merit, improving the background rejection, at 30\% signal efficiency, by 15 - 20\%. The use of attention coefficients allow the model complexity of ABCNet to be reduced, having 40\% less parameters compared to ParticleNet.

\begin{table}[h]
    \centering
    \caption{Comparison between the performance achieved with ABCNet and different available implementations. The uncertainty quoted corresponds to the standard deviation of nine trainings with different random weight initialisation. If the uncertainty is not quoted then the variation is negligible compared to the expected value.}
    \label{tab:qg_results}
	\begin{tabular}{lccccc}
    \noalign{\smallskip}\hline
          &  Acc &AUC & 1/$\epsilon_B$ ($\epsilon_S = 0.5$)  & 1/$\epsilon_B$ ($\epsilon_S = 0.3$) & Parameters\\
            \hline
            ResNeXt-50 & 0.821 & 0.9060 & 30.9 & 80.8 & 1.46M\\
            P-CNN & 0.827 & 0.9002 & 34.7 & 91.0 & 348k \\
            PFN & - & 0.9005 & 34.7$\pm$0.4 & - & 82k \\
            ParticleNet-Lite & 0.835 & 0.9079 &  37.1 & 94.5 & \textbf{26k} \\
            ParticleNet & \textbf{0.840} & 0.9116 & 39.8$\pm$0.2 & 98.6$\pm$1.3 & 366k\\
            ABCNet & \textbf{0.840} & \textbf{0.9126} & \textbf{42.6$\pm$0.4} & \textbf{118.4$\pm$1.5} & 230k\\
	\noalign{\smallskip}\hline    
	\end{tabular}
\end{table}

\subsection{Visualisation}
\begin{figure}[ht]
    \centering
    \resizebox{0.75\textwidth}{!}{
        \includegraphics{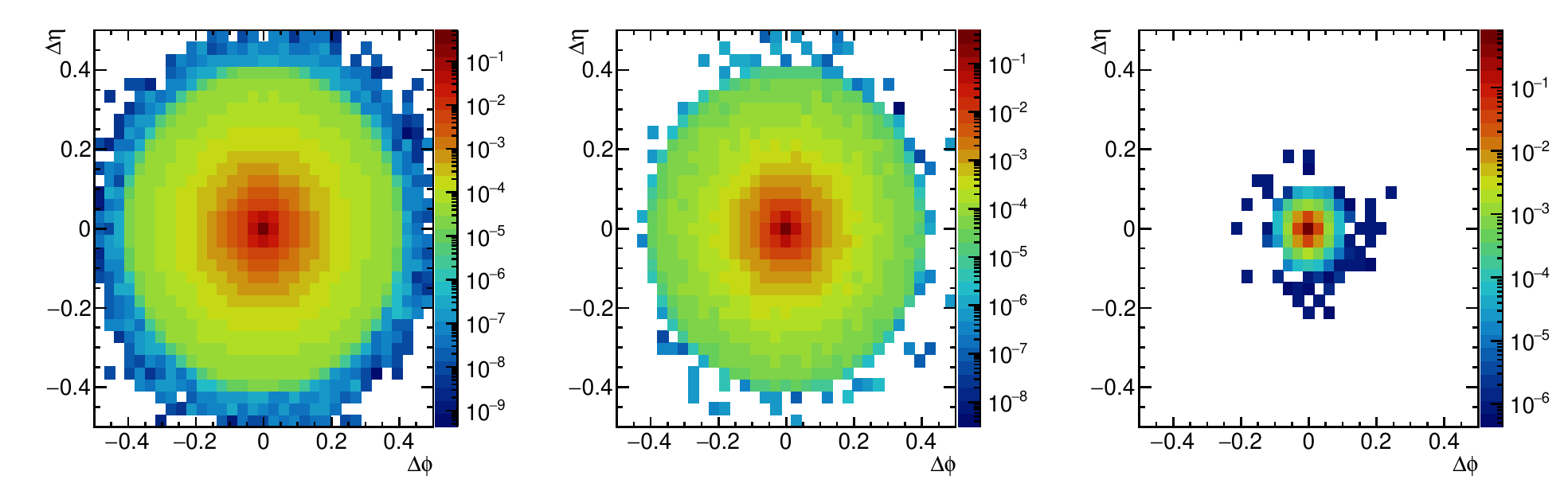}
    }   
    \resizebox{0.75\textwidth}{!}{
        \includegraphics{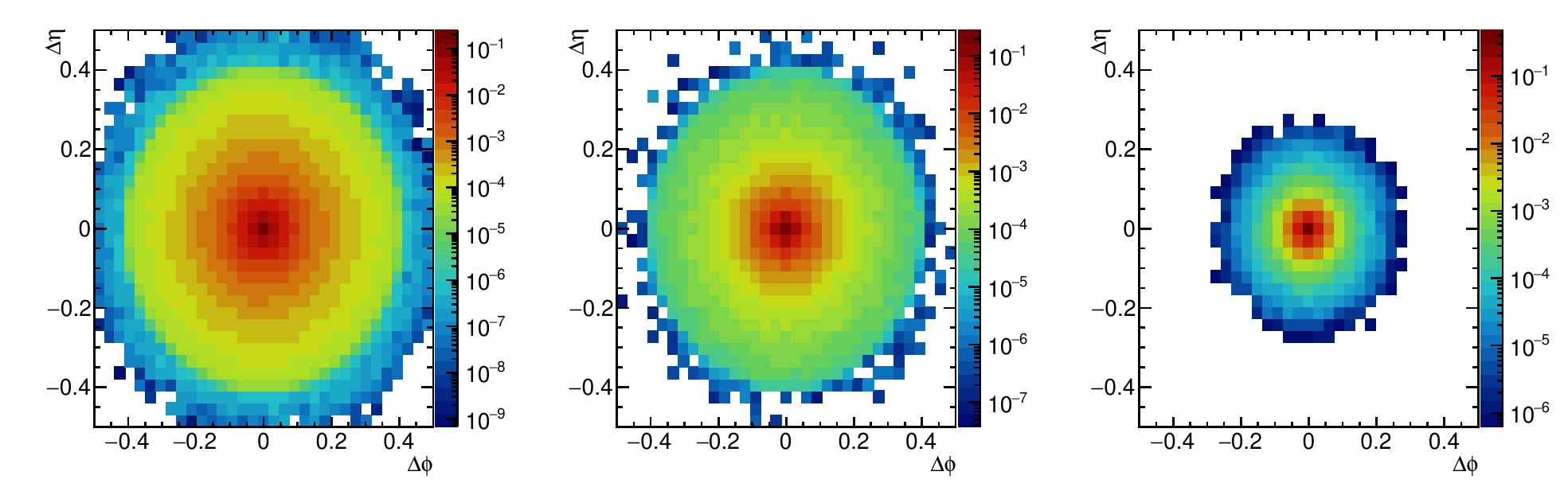}   
    }   
            
    \caption{Distribution of the $\pt$-scaled distribution of the jet constituents averaged over all images in the test sample. The leftmost images are the quark (top) and gluon (bottom) jet averages after the pre-processing. The first 5$\%$ of the jet constituents with the highest self-attention coefficients for the first and second GAPLayers are shown on the images in the centre and right, respectively.}
    \label{fig:att_sig}
\end{figure}

A simple way to check what ABCNet is learning is to look at the self-coefficients of each point of the point cloud. First, we pre-processes the images  in a similar fashion as \cite{Komiske:2016rsd}, using the following steps:
\begin{itemize}
    \item Centre: All jet images are translated in the $\eta-\phi$ space to a common centre at (0,0). The centre of the jet is taken as its $\pt$-weighted centroid.
    \item Particle scale: Each particle constituent has its transverse momentum scaled such that $\sum^{jet}p_{\mathrm{T,i}} = 1$, where $i$ is the i-th constituent of the jet.
    \item Overall scale: The final image is created by superimposing the individual event images and  dividing the resulting distribution by the number of events in the test sample.
\end{itemize}
Other steps were adopted in \cite{Komiske:2016rsd}, however since the goal is to have a simple visual cue, they were not used.
The resulting jet images are shown in Fig.~\ref{fig:att_sig} for quark- and gluon-initiated jets on the upper and lower rows, respectively. The leftmost images correspond to the jets after the pre-processing. The subsequent columns show the same distribution, but only considering particles whose self-attention coefficients, resulting from the first (middle column) and second (right column) GAPLayers, are higher than a certain value. This value is chosen such that only the first 5$\%$ of all particles with the largest self-attention coefficients are selected. The self-coefficients from the first GAPLayer have the effect of giving higher attention to high-$\pt$ particles while soft-QCD with large angular variation has less importance. The second GAPLayer, where nearest-neighbours are calculated in the feature space, have different distributions for quark-initiated and gluon-initiated jets. Quark initiated jets have the highest coefficients in a confined radius with $\sim \Delta R = 0.1$ around the centre, while gluon initiated coefficients spam a bigger area around the centre with $\sim \Delta R = 0.3$. That behaviour is expected since gluon-jets have a larger colour factor compared to quark jets, typically resulting in a broader angular distribution compared to quark jets.

\section{Pileup reduction using part segmentation}
Another crucial problem in particle physics is how to identify the particles originated from high-$\pt$ collisions, and separate them from unwanted additional interactions. Two traditional methods to accomplish this task are the Softkiller \cite{Cacciari:2014gra} and the Pileup Per Particle Identification (PUPPI) \cite{Bertolini:2014bba} algorithms. These two algorithms are chosen since they  represent the most common algorithms for pileup mitigation at the LHC. To test the performance of ABCNet in this context, we change the scope of a single jet classifier to a particle-by-particle classification (part segmentation). In this case, a probability is estimated for object, determining how likely it is for each particle to originate from the leading vertex (LV). The sample used for this study is available from \cite{Komiske:2017ubm}, containing a set of $q\bar q$ light-quark-initiated jets coming from the decay of a scalar particle with mass $m_\phi = $ 500 GeV. The samples were generated using  \textsc{Pythia8} at $\sqrt{s}$ = 13 TeV. The pileup events were generated by overlaying soft QCD processes onto each event. Stable particles are clustered into jets, excluding neutrinos, using the anti-$k_{T}$ algorithm with R=0.4. At parton level, a $\pt$ requirement of at least 95 GeV was applied. Only jets satisfying $\pt$>100 GeV and $\eta\in$[-2.5,2.5] are considered. For each event, up to two leading jets as ordered in $\pt$ are stored. Two thousand events are generated, each with a different number of pileup interactions (NPU) ranging from 0 to 180. For the training and testing samples, events are randomly selected from the generated samples according to a Poisson distribution with average pileup <NPU> = 140, motivated by the expected pileup levels for future collisions at the LHC. The training and evaluation are done with 80\% and 10\% of the events with <NPU> = 140, respectively. For testing, two samples are created: one corresponding to the remaining 10\% of the events and  <NPU> = 140 and the other a sample with independent events generated at different NPU levels. For each event, up to 500 particles are stored as long as they are matched to one of the two leading jets. The features used to define each particle are described in Table~ \ref{tab:pu_vars}. The feature choice is similar to the ones used for the classification task. The main difference is that for this sample, the PID information is not available, but replaced by a flag that identifies if a particle is charged or not. Since more than one jet can be reconstructed, a global zero is used for all events, instead of using the jet axis as a reference point. While no selection is applied to the particles used in ABCNet, the PUPPI weights and the SoftKiller decision flag  are also used as input features. The global information added to the parameterisation is NPU and the number of reconstructed particles associated to jets.

\begin{table}[h]
    \centering
   
	\begin{tabular}{lc}
	\noalign{\smallskip}\hline
             Variable & Description  \\
            \hline
            $\eta$       &  \small{Particle's pseudo-rapidity.}\\  
            $\phi$       &  \small{Particle's azimuthal angle.}\\  
            $\log\pt$       &  \small{logarithm of the particle $\pt$. }\\  
            $Q$       &  \small{boolean flag identifying if the particle is charged.}\\  
            $\log\frac{\pt}{\pt(\mathrm{jet})}$       &  \small{logarithm of the ratio between the particle $\pt$ and the  associated jet $\pt$.}\\  
            $\log\frac{\mathrm{E}}{\mathrm{E}(\mathrm{jet})}$       &  \small{logarithm of the ratio between the particle E and the associated jet E.}\\  
            $w_{\mathrm{PUPPI}}$       &  \small{PUPPI weight for the particle.}\\ 
            $w_\mathrm{SoftKiller}$      &  \small{boolean flag identifying if the particle passes the SoftKiller $\pt$ requirement.}\\
            \hline
            NPU & \small{number of pileup interactions.} \\
            NPART & \small{number of reconstructed particles associated to jets.}  \\
    \noalign{\smallskip}\hline
	\end{tabular}
	\caption{Variable description for each feature used to define a point in the point cloud implementation for the pileup mitigation problem. The latter two features are the global information added to the network.}
	 \label{tab:pu_vars}
\end{table}

\subsection{Network architecture}
\begin{figure}[ht]
    \centering
    \resizebox{0.75\textwidth}{!}{
        \includegraphics{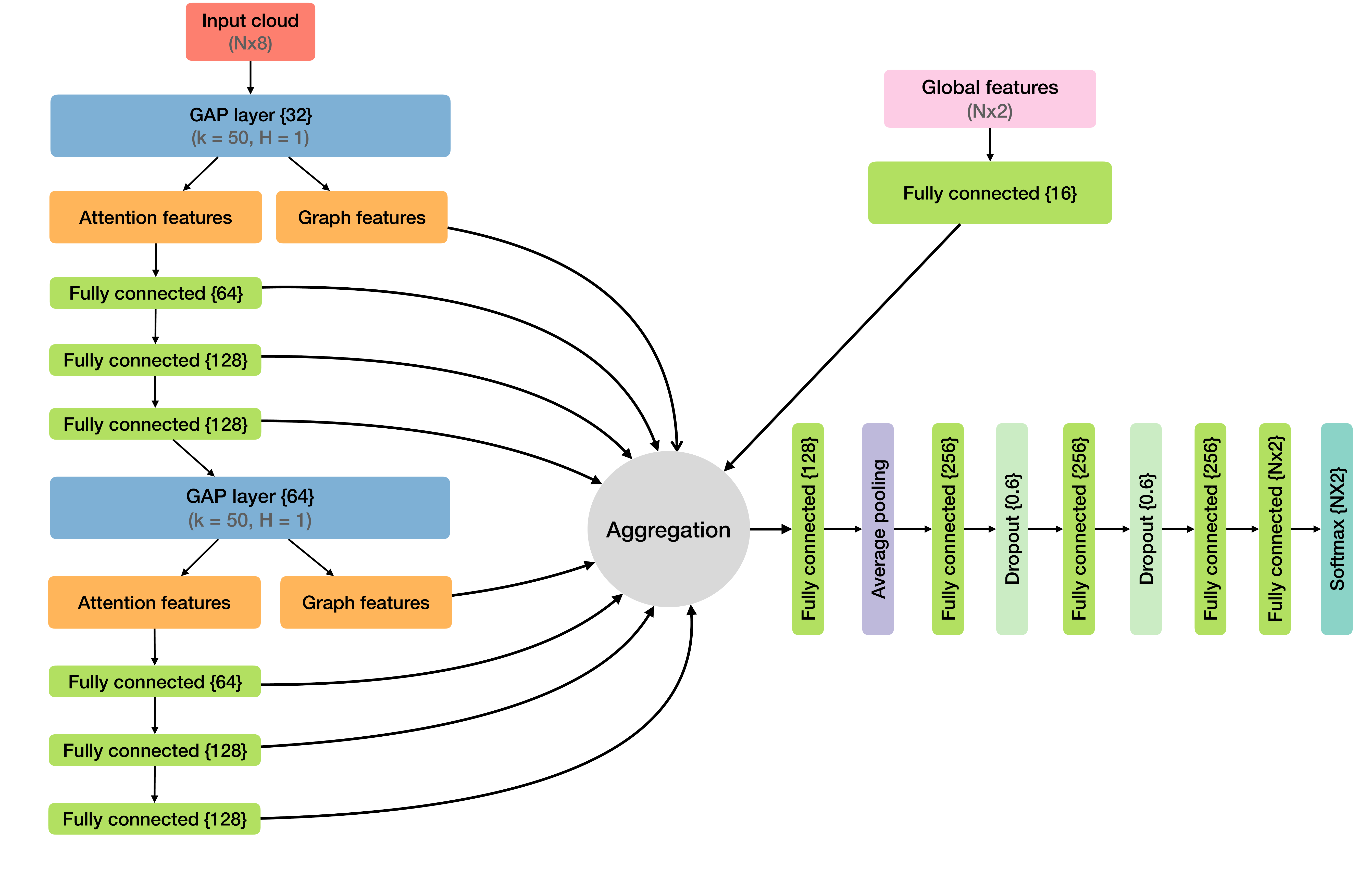}
    } 
    
    \caption{ABCNet architecture used for pileup identification. Fully connected layer and encoding node sizes are denoted inside ``\{\}''. For each GAPLayer, the number of k-nearest neighbours (k) and heads (H) are given.}
    \label{fig:ABC_PU}
\end{figure}

The network architecture for the part segmentation problem is similar to the setup used previously. The main differences are:
\begin{itemize}
    \item Number of considered neighbours increased from 10 to 50.
    \item Additional MLPs after the attention features and after the pooling layer.
    \item Usage of only 1-head GAPLayers.
\end{itemize}
The increase in expansion of neighbours and MLPs are chosen to increase the model's capacity to cover the larger amount of points per event. 
The architecture is shown in Fig.~\ref{fig:ABC_PU}.

\subsection{Results}
\begin{figure}[ht]
    \centering
     \resizebox{0.75\textwidth}{!}{
        \includegraphics{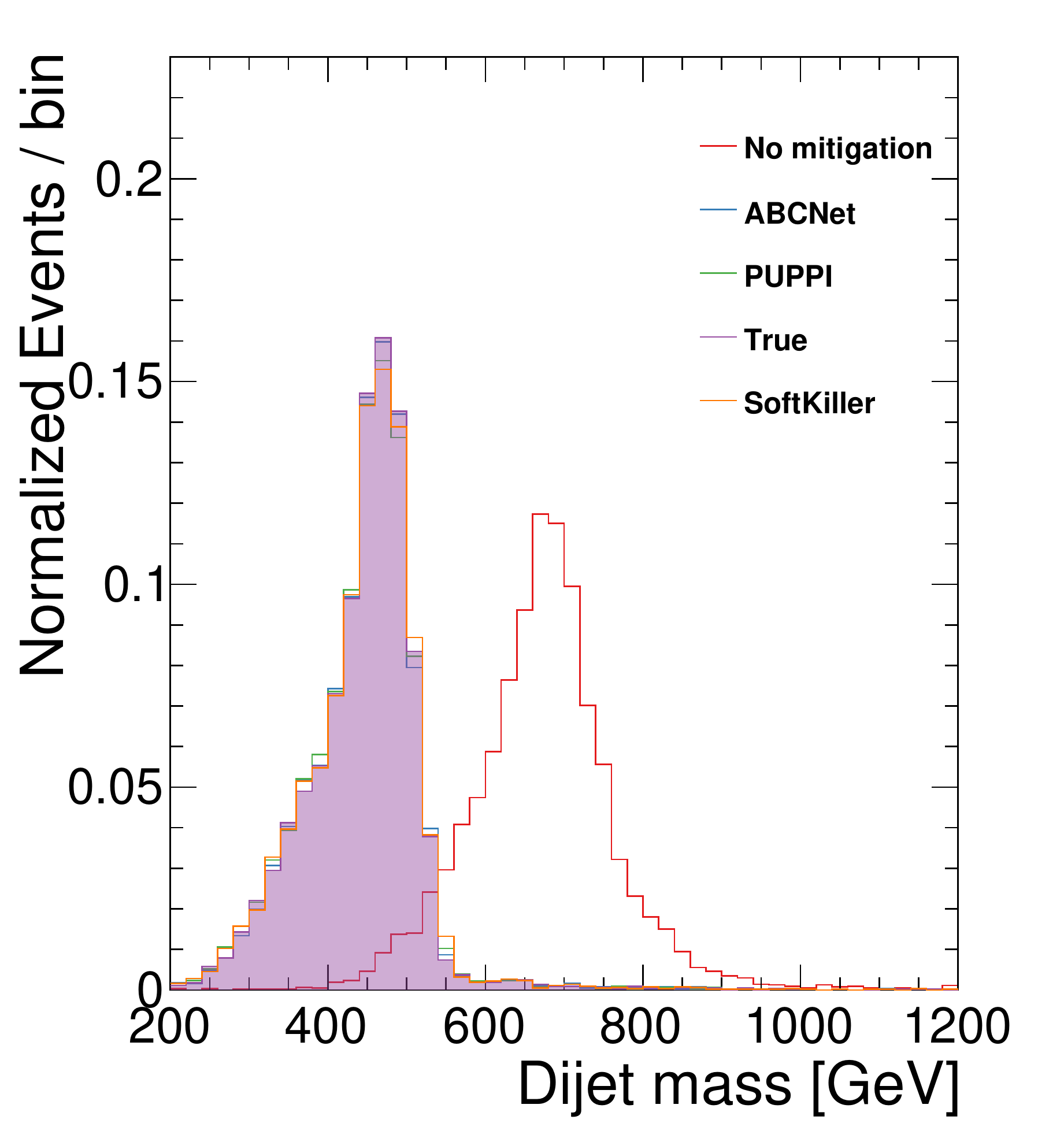}
        \includegraphics{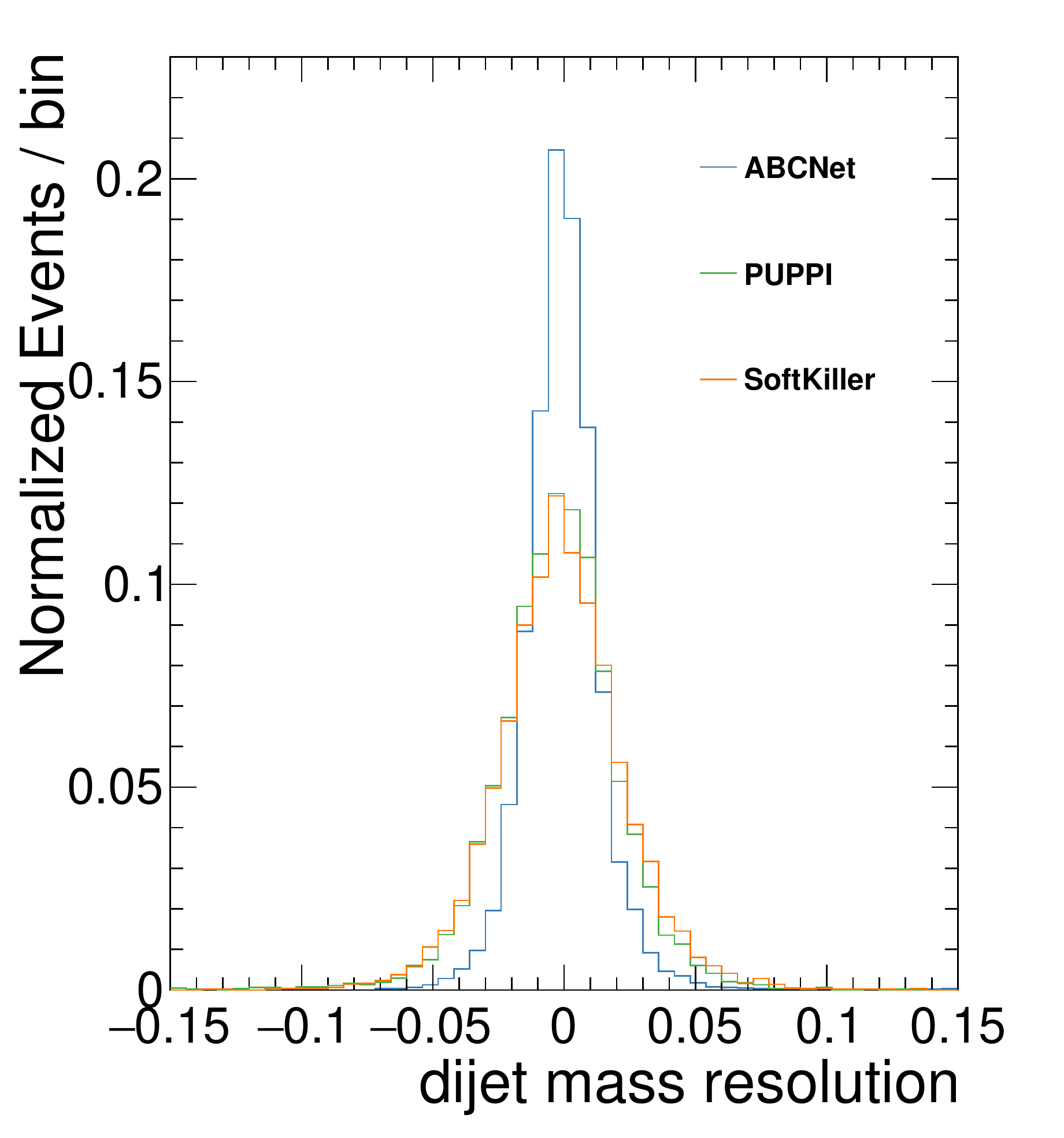}
    }

    \caption{Distribution of the dijet mass using the different pileup mitigation algorithms (left) and the jet mass resolution (right). A narrower resolution peak means better performance. All distributions are normalised to unit.}
    \label{fig:dijetmass}
\end{figure}

The performance of ABCNet is compared to the performance achieved using PUPPI and SoftKiller. 
The default parameters for those methods are the same as the ones used in \cite{Komiske:2017ubm}: $R_0=$0.3, $R_{min}$ = 0.02, $w_{cut}$ = 0.1, $\pt^{cut}$(NPU) = 0.1 + 0.007 $\times$ NPU (PUPPI), grid size = 0.4 (SoftKiller). First, the jet mass is reconstructed with the <NPU>=140 evaluation sample, applying the different mitigation algorithms. Inspired by PUPPI, the output probabilities from ABCNet are used to reweight the four-momentum of each particle. The reconstructed dijet mass and the dijet mass resolution are shown in Fig.~\ref{fig:dijetmass}. The resolution is defined as:
\begin{linenomath}
\begin{equation*}
    \mathrm{mass ~ resolution} = \frac{m_{reco}-m_{true}}{m_{true}}.
\end{equation*}
\end{linenomath}
In Table~\ref{tab:pu_res}, the width of the jet mass resolution, extracted by fitting the distributions in Fig.~\ref{fig:dijetmass} (right) with a Gaussian function, is also listed.

\begin{table}[h]
    \centering
	\begin{tabular}{lc}
	\noalign{\smallskip}\hline
            \hline
            Algorithm  &  Resolution width \\
            \hline
            SoftKiller   &  0.022\\  
            PUPPI   &  0.021\\  
            ABCNet   &  \textbf{0.012}\\  
    \noalign{\smallskip}\hline
	\end{tabular}
    \caption{Resolution width for different pileup mitigation strategies. The resolution width is extracted by fitting the distributions shown in Fig.~\ref{fig:dijetmass} (right) with a Gaussian function.}
    \label{tab:pu_res}
\end{table}

\begin{figure}[ht]
    \centering 
    \resizebox{0.75\textwidth}{!}{
        \includegraphics{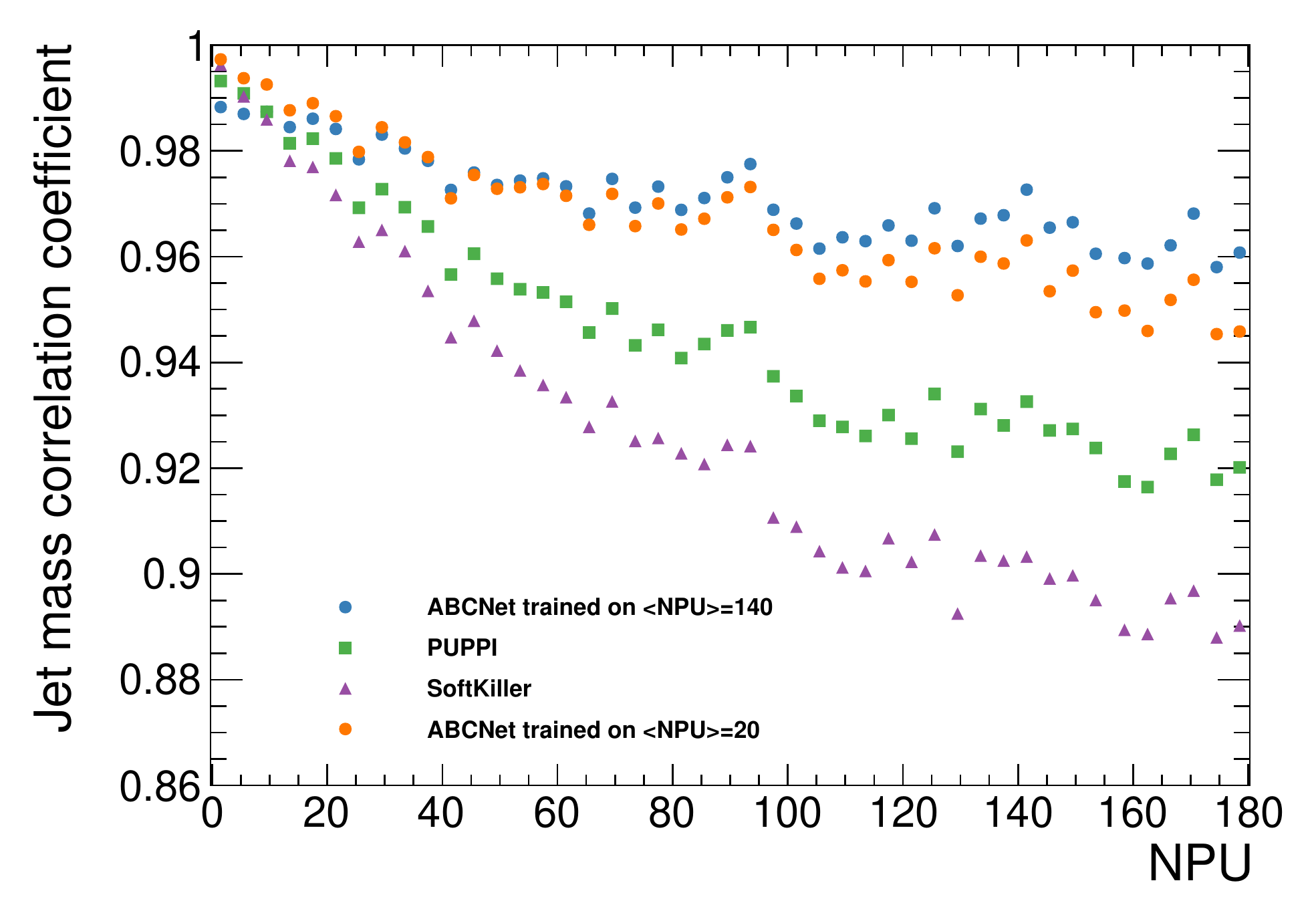}
    }

    \caption{PCC for each pileup mitigation algorithm for different NPU. ABCNet is trained on <NPU>=140 (blue) or <NPU>=20 (orange).}
    \label{fig:corr}
\end{figure}

ABCNet improves jet mass resolution compared to both PUPPI and SoftKiller by 75\% and 83\%, respectively.
The robustness of each algorithm is also tested by comparing The Pearson linear correlation coefficient (PCC) between the true jet mass and corrected jet masses for different NPU is generated. Figure \ref{fig:corr} shows the result of the comparison using the test sample with NPU from 0 to 180. To investigate the power of ABCNet to generalise, a training sample with <NPU> = 20 is created and trained using the same architecture described previously. For both trainings, ABCNet shows a superior performance compared to PUPPI and SoftKiller for the entire NPU range. Furthermore, ABCNet is also remarkably robust for pileup variations outside the training region due to the addition of the global parameters to the method.

\section{Training details}
ABCNet is implemented using Tensorflow v1.4 \cite{tensorflow2015-whitepaper}. A Nvidia GTX 1080 Ti graphics card is used for the training and evaluation steps. For all tasks described in this paper, the Adam optimiser \cite{2014arXiv1412.6980K} is used. The learning rate starts from 0.001 and decreases linearly by a factor 10 every seven epochs, until reaching a minimum of 1e-7. The training is performed with a mini batch size of 64 to a maximum number of 50 epochs. The epoch with the highest accuracy on the evaluation is saved in the case of the quark-gluon classification task. For the pileup identification, the epoch with the lowest loss is stored.  

\section{Conclusion}
In this document, a new machine learning implementation for data classification in HEP is introduced. The attention-based cloud net (ABCNet) takes advantage of the data structure commonly found in particle colliders to create a point cloud interpretation. An attention mechanism is implemented to enhance the local information extraction and provide a simple way to investigate what the method is learning. To capture the global information, direct connections for global input features can be directly added. ABCNet can be used for event-by-event classification problems or generalised to particle-by-particle classification. To exemplify the architecture flexibility, two example problems are investigated: quark-gluon classification and pileup mitigation. For both problems, ABCNet achieved an improved performance compared to other available methods. By using a graph architecture and interpreting each point in a point cloud as a particle, ABCNet can be readily adapted to other applications in HEP like jet-flavour tagging, boosted jet identification, or particle-track reconstruction. 

\section{Acknowledgements}
This research was supported in part by the Swiss National Science Foundation (SNF) under contract No. 200020-182037. The authors would like to thank Loukas Gouskos and Ben Kilminster for the valuable suggestions regarding the development and clarity of this document.

%
%
%

%
\bibliographystyle{unsrt}
\bibliography{ref}

\begin{thebibliography}{10}

\bibitem{Evans:2008zzb}
Lyndon Evans and Philip Bryant.
\newblock {LHC Machine}.
\newblock {\em JINST}, 3:S08001, 2008.

\bibitem{Aad:2008zzm}
ATLAS Collaboration.
\newblock {The ATLAS Experiment at the CERN Large Hadron Collider}.
\newblock {\em JINST}, 3:S08003, 2008.

\bibitem{Chatrchyan:2008aa}
{CMS Collaboration}.
\newblock {The CMS Experiment at the CERN LHC}.
\newblock {\em JINST}, 3:S08004, 2008.

\bibitem{2019arXiv190508705C}
Can {Chen}, Luca {Zanotti Fragonara}, and Antonios {Tsourdos}.
\newblock {GAPNet: Graph Attention based Point Neural Network for Exploiting
  Local Feature of Point Cloud}.
\newblock {\em arXiv e-prints}, page arXiv:1905.08705, May 2019.

\bibitem{2019arXiv190208570Q}
Huilin {Qu} and Loukas {Gouskos}.
\newblock {ParticleNet: Jet Tagging via Particle Clouds}.
\newblock {\em arXiv e-prints}, page arXiv:1902.08570, February 2019.

\bibitem{NIPS2017_6931}
Manzil Zaheer, Satwik Kottur, Siamak Ravanbakhsh, Barnabas Poczos, Ruslan~R
  Salakhutdinov, and Alexander~J Smola.
\newblock Deep sets.
\newblock In I.~Guyon, U.~V. Luxburg, S.~Bengio, H.~Wallach, R.~Fergus,
  S.~Vishwanathan, and R.~Garnett, editors, {\em Advances in Neural Information
  Processing Systems 30}, pages 3391--3401. Curran Associates, Inc., 2017.

\bibitem{Komiske_2019}
Patrick~T. Komiske, Eric~M. Metodiev, and Jesse Thaler.
\newblock Energy flow networks: deep sets for particle jets.
\newblock {\em Journal of High Energy Physics}, 2019(1), Jan 2019.

\bibitem{bib:message_passing}
A.~Lister J.~Pearkes S.~Egan, W.~Fedorko and C.~Gay.
\newblock {Neural Message Passing for Jet Physics}.
\newblock In {\em Proceedings of the Deep Learning for Physical Sciences
  Workshop at NIPS (2017)}, 2017.

\bibitem{Moreno:2019bmu}
Eric~A. Moreno, Olmo Cerri, Javier~M. Duarte, Harvey~B. Newman, Thong~Q.
  Nguyen, Avikar Periwal, Maurizio Pierini, Aidana Serikova, Maria Spiropulu,
  and Jean-Roch Vlimant.
\newblock {JEDI-net: a jet identification algorithm based on interaction
  networks}.
\newblock 2019.

\bibitem{Abdughani:2018wrw}
Murat Abdughani, Jie Ren, Lei Wu, and Jin~Min Yang.
\newblock {Probing stop pair production at the LHC with graph neural networks}.
\newblock {\em JHEP}, 08:055, 2019.

\bibitem{DBLP:journals/corr/abs-1809-06166}
Nicholas Choma, Federico Monti, Lisa Gerhardt, Tomasz Palczewski, Zahra
  Ronaghi, Prabhat, Wahid Bhimji, Michael~M. Bronstein, Spencer~R. Klein, and
  Joan Bruna.
\newblock Graph neural networks for icecube signal classification.
\newblock {\em CoRR}, abs/1809.06166, 2018.

\bibitem{osti_1484458}
Steven Farrell et~al.
\newblock {Novel deep learning methods for track reconstruction}.
\newblock In {\em {4th International Workshop Connecting The Dots 2018
  (CTD2018) Seattle, Washington, USA, March 20-22, 2018}}, 2018.

\bibitem{Qasim2019}
Shah~Rukh Qasim, Jan Kieseler, Yutaro Iiyama, and Maurizio Pierini.
\newblock Learning representations of irregular particle-detector geometry with
  distance-weighted graph networks.
\newblock {\em The European Physical Journal C}, 79(7):608, Jul 2019.

\bibitem{ArjonaMartinez2019}
J.~Arjona~Mart{\'i}nez, O.~Cerri, M.~Spiropulu, J.~R. Vlimant, and M.~Pierini.
\newblock Pileup mitigation at the large hadron collider with graph neural
  networks.
\newblock {\em The European Physical Journal Plus}, 134(7):333, Jul 2019.

\bibitem{DBLP:journals/corr/abs-1801-07829}
Yue Wang, Yongbin Sun, Ziwei Liu, Sanjay~E. Sarma, Michael~M. Bronstein, and
  Justin~M. Solomon.
\newblock Dynamic graph {CNN} for learning on point clouds.
\newblock {\em CoRR}, abs/1801.07829, 2018.

\bibitem{velikovi2017graph}
Petar Velickovic, Guillem Cucurull, Arantxa Casanova, Adriana Romero, Pietro
  Lio, and Yoshua Bengio.
\newblock Graph attention networks, 2017.

\bibitem{Cacciari:2008gp}
Matteo Cacciari, Gavin~P. Salam, and Gregory Soyez.
\newblock The anti-$k_{T}$ jet clustering algorithm.
\newblock {\em JHEP}, 04:063, 2008.

\bibitem{Sjostrand:2014zea}
Torbjörn Sj{\"o}strand, Stefan Ask, Jesper~R. Christiansen, Richard Corke,
  Nishita Desai, Philip Ilten, Stephen Mrenna, Stefan Prestel, Christine~O.
  Rasmussen, and Peter~Z. Skands.
\newblock {An Introduction to PYTHIA 8.2}.
\newblock {\em Comput. Phys. Commun.}, 191:159--177, 2015.

\bibitem{Baldi:2016fzo}
Pierre Baldi, Kyle Cranmer, Taylor Faucett, Peter Sadowski, and Daniel
  Whiteson.
\newblock {Parameterized neural networks for high-energy physics}.
\newblock {\em Eur. Phys. J.}, C76(5):235, 2016.

\bibitem{Tanabashi:2018oca}
M.~Tanabashi et~al.
\newblock {Review of Particle Physics}.
\newblock {\em Phys. Rev.}, D98(3):030001, 2018.

\bibitem{Komiske:2016rsd}
Patrick~T. Komiske, Eric~M. Metodiev, and Matthew~D. Schwartz.
\newblock {Deep learning in color: towards automated quark/gluon jet
  discrimination}.
\newblock {\em JHEP}, 01:110, 2017.

\bibitem{Cacciari:2014gra}
Matteo Cacciari, Gavin~P. Salam, and Gregory Soyez.
\newblock {SoftKiller, a particle-level pileup removal method}.
\newblock {\em Eur. Phys. J.}, C75(2):59, 2015.

\bibitem{Bertolini:2014bba}
Daniele Bertolini, Philip Harris, Matthew Low, and Nhan Tran.
\newblock {Pileup Per Particle Identification}.
\newblock {\em JHEP}, 10:059, 2014.

\bibitem{Komiske:2017ubm}
Patrick~T. Komiske, Eric~M. Metodiev, Benjamin Nachman, and Matthew~D.
  Schwartz.
\newblock {Pileup Mitigation with Machine Learning (PUMML)}.
\newblock {\em JHEP}, 12:051, 2017.

\bibitem{tensorflow2015-whitepaper}
{M. Abadi et al.}
\newblock {TensorFlow}: Large-scale machine learning on heterogeneous systems,
  2015.
\newblock Software available from tensorflow.org.

\bibitem{2014arXiv1412.6980K}
Diederik~P. {Kingma} and Jimmy {Ba}.
\newblock {Adam: A Method for Stochastic Optimization}.
\newblock {\em arXiv e-prints}, page arXiv:1412.6980, Dec 2014.

\end{thebibliography}

\end{document}